\pdfoutput=1
\documentclass[11pt,oneside,british,reqno]{article}                  
\usepackage{latexsym}  
\usepackage{amsmath,amssymb}
\usepackage{graphicx}                 
\usepackage{lmodern}
\usepackage{cite}
\usepackage[T1]{fontenc}
\usepackage[latin9]{inputenc}
\usepackage[a4paper]{geometry}
\usepackage{color}
\usepackage{babel}
\usepackage{pifont}
\usepackage{amssymb}
\usepackage{fixltx2e}
\usepackage{setspace}
\usepackage{hyperref}

\numberwithin{equation}{section}
\numberwithin{figure}{section}
\numberwithin{table}{section}

\renewcommand\[{\begin{equation}}
\renewcommand\]{\end{equation}}
\usepackage{titlesec}
\usepackage{dsfont}
\usepackage {url}
\usepackage{slashed}
\usepackage{yfonts}

\begin{document}

\title{Tropical Limit in Statistical Physics}

\author{M. Angelelli and B. Konopelchenko,\\Department of Mathematics and Physics ``Ennio De Giorgi'',\\ 
University of Salento and sezione INFN, \\
Lecce, 73100, Italy.}
\date{}
\maketitle

 \abstract
Tropical limit for macroscopic systems in equilibrium defined as the
formal limit of Boltzmann constant $k\rightarrow0$ is discussed.
It is shown that such tropical limit is well-adapted to analyse properties
of systems with highly degenerated energy levels, particularly of
frustrated systems like spin ice and spin glasses. Tropical free energy
$F_{tr}(T)$ is a piecewise linear function of temperature $T$, tropical entropy is a piecewise constant function and the system has energy for which tropical Gibbs' probability has maximum.
Properties of systems in the points of jump of entropy are studied.
Systems with finite and infinitely many energy levels and phenomena
of limiting temperatures are discussed.

\section{Introduction}

Singular (nonanalytic) limits of various types have shown up many
times in physics and mathematics. Maslov's dequantization \cite{1,2,3},
ultra-discrete integrable systems \cite{4,5,6,7,8}
and tropical geometry \cite{9,10,11,12,13} are
three apparently disconnected fields where such a limit was most actively
studied during last twenty years. Nowadays all of them are viewed
as the different faces of the so-called tropical mathematics (see
e.g. \cite{14,15,16}). Tropical limit is
characterized by a highly singular limiting behavior of the type ${\displaystyle x=\exp\left(\frac{x_{tr}}{\varepsilon}\right)}$
as the parameter $\varepsilon\rightarrow0$. Elements $x_{tr}$ form
an idempotent semiring with the tropical addition $\oplus$ and multiplication
$\odot$ defined by $x_{1tr}\oplus x_{2tr}={\displaystyle  \lim_{\varepsilon\rightarrow0} \left(\varepsilon \ln \left( \exp\frac{x_{1tr}}{\varepsilon}+\exp\frac{x_{2tr}}{\varepsilon}\right)\right)}$$=\max\{x_{1tr},x_{2tr}\}$
and $x_{1tr}\odot x_{2tr}={\displaystyle \lim_{\varepsilon\rightarrow0}  \left(\varepsilon \ln \left( \exp\frac{x_{1tr}}{\varepsilon}\cdot\exp\frac{x_{2tr}}{\varepsilon}\right)\right)=x_{1tr}+x_{2tr}}$ \cite{9,10,11,12,13,14,15,16}. 

It was already noted in \cite{13,17,18,19,20,21} that
statistical physics seems to be the part of physics most naturally
adapted to consider the tropical limit. Indeed, free energy $F$
of the macroscopic system in equilibrium is given by the formula \cite{22}  

\begin{equation}
{\displaystyle F=-kT\ln{\displaystyle \sum_{n}g_{n}\exp\left(-\frac{E_{n}}{kT}\right)}}\label{eq: free energy}
\end{equation}
where $k$ is the Boltzmann constant, $T$ is the absolute temperature,
$\{E_{n}\}$ is the energy spectrum of the system, $g_{n}$ are\textcolor{red}{{}
}statistical weights (degeneracies) of the corresponding levels $E_{n}$ 
and the sum is performed over different energy levels. Thus, in the
limit $kT\rightarrow0$ one has the tropical sum in the r.h.s. of
the formula (\ref{eq: free energy}) and $E_{n}$ and $F(kT\rightarrow0)$
become elements of idempotent semiring refered in \cite{21}
as the thermodynamic semiring. In the papers \cite{13,19,20,21}
the tropical limit was identified with the limit $T\rightarrow0$.
With such a choice tropical free energy is equal to $E_{min}$ and entropy $S_{tr}=0$
for the systems with finite $g_{n}$. 

In this paper we argue that the formal limit $k\rightarrow0$ is more
appropriate avatar of tropical limit in statistical physics. At first
glance the separation of $k$ and $T$ seems to be artificial and
irrelevant since the r.h.s. of (\ref{eq: free energy}) and Gibbs'
distribution 

\begin{equation}
{\displaystyle w_{n}=\frac{{\displaystyle \exp \left(-\frac{E_{n}}{kT}\right)}}{{\displaystyle \sum_{m}g_{m}\exp\left(-\frac{E_{m}}{kT}\right)}}}\label{eq: Gibbs' distribution}
\end{equation}
contain only the product $kT$. It is indeed so for systems with
finite $g_{n}$. 

An observation is that there exists a wide class of systems with exponentially
large degeneracies $g_{n}$ for which the situation is quite different.
At 1935 L. Pauling \cite{23} showed
that the degeneracy of the ground state of the ice is given by $g_{0}={\displaystyle \exp(N \ln\frac{3}{2})}$,
where $N$ is the number of molecules. So the ice has (residual) entropy
$S_{0}={\displaystyle kN \ln\frac{3}{2}}$ at $T=0$ that is in excellent
agreement with experimental data \cite{24}. 
Several other systems like spin ices and spin glasses have exponentially
large degeneracies of ground and excited states of the type $g_{n}={\displaystyle \exp(a_{n}N)}$
with certain constants $a_{n}$ (see e.g. \cite{22,25,26,27,28,29,30,31,32,33}). 
In the thermodynamic limit $N\rightarrow\infty$ such $g_{n}$ have
typical tropical behavior. A natural way to formalize this limit
is to represent exponentially large degeneracies as $g_{n}=\exp{\displaystyle \frac{S_{n}}{k}}$
with finite $S_{n}$ and $k\rightarrow0$. 
Physically it corresponds to the limit $N\rightarrow\infty$, $k\rightarrow0$
with $k\cdot N=$constant (gas constant $R$) and $\displaystyle S_{n}=a_{n} R$.

Thus, representing the degeneracies $g_{n}$ as $g_{n}=\exp{\displaystyle \frac{S_{n}}{k}}$
and defining $F_{tr}={\displaystyle \lim_{k\rightarrow0}F}$, one
has at $T>0$ 

\begin{equation}
{\displaystyle F_{tr}(T)=-T\sum_{n}\oplus\left(-\frac{F_{n}}{T}\right)=\min\{F_{1},F_{2},..,F_{n},..\}}\label{eq: tropical free energy}
\end{equation}
where $F_{n}=E_{n}-TS_{n}$ is a ``microscopic'' free energy associated
with the energy level $E_{n}$. So $F_{tr}(T)$ is a piecewise linear
function of temperature $T$. 
This leads to various consequences. For instance, the tropical entropy
$S_{tr}={\displaystyle -\frac{\partial F_{tr}}{\partial T}}$$=S_{n_{min}}$
where $n_{min}$ is the index of minimal free energy $F_{n_{min}}$ at temperature $T$ in the case when the minimum is attained only once.
So $S_{tr}$ is a piecewise constant function of $T$. The value $S_{tr}(T=0)$
is the residual entropy of the macroscopic system at $T=0$. At certain singular
values of $T$ $S_{tr}$ exhibits jumps (entropy drop). Depending on the system it happens either at positive or negative temperatures. 

These properties of the tropical limit $k\rightarrow0$
trace quite well certain characteristic features of various frustrated
systems similar to spin ices and spin glasses. In constrast these
properties get lost in the limit $T\rightarrow0$. 

This is the main evidence in favour of the definition of the tropical
limit as $k\rightarrow0$. The second reason is that in such a limit
the basic thermodynamic equations, like the first law $dE=TdS-pdV$
and relations between thermodynamic potentials, remain unaltered leaving
temperature $T$ to be a free positive or negative parameter. In addition
the limit $k\rightarrow0$ resembles very much that of $\hbar\rightarrow0$
in Maslov's dequantization. 

Tropical limit of Gibbs' distribution (\ref{eq: Gibbs' distribution})
has rather interesting properties too. Tropical probability $w_{n,tr}={\displaystyle \lim_{k\rightarrow0}(k\cdot\ln w_{n})}$
takes values in the interval $(-\infty,0]$ and is equal to 

\begin{equation}
{\displaystyle w_{n,tr}=-S_{n}+\frac{F_{tr}-F_{n}}{T}}.\label{eq: tropical Gibbs' distribution}
\end{equation}
The tropical probability $W_{n,tr}$ for the system to have energy $E_{n}$ is 
\begin{equation}
{\displaystyle{W_{n,tr}=w_{n,tr}+S_{n}=\frac{F_{tr}-F_{n}}{T}}}\label{eq: tropical Gibbs' distribution for levels}
\end{equation}
 and it is normalized by the condition $\displaystyle{\sum_{n} \oplus W_{n,tr}=0}$.

These tropical Gibbs' distributions
describe fine structure of the states with exponentially small usual
probabilities $w_{n}\sim{\displaystyle \exp\left(-\frac{S_{n}}{k}\right)}$. It
is shown that tropical probabilities and entropy have a peculiar behavior at
the singular values $T^{*}$ of temperature at which jump of $S_{tr}$
is observed. 

Systems with finitely many energy levels are considered as illustrative
examples. Tropical limit of the systems with infinite number of energy
levels, the phenomenon of limiting temperatures and existence of intervals
of forbidden temperatures are discussed too. 

It is noted that the
limit $k\rightarrow0$ viewed as the limit of vanishing white noise
for systems with finite degeneracies has been discussed in a different
context in \cite{34}. 

The paper is organized as follows. In section $2$ general definitions
and formulas are presented. Singularities appearing in tropical limit
are analysed in next section $3$. Systems with finite number of energy
levels are considered in section $4$. In section $5$ the systems
with infinitely many energy levels bounded and unbounded from below
and the existence of limiting temperatures are discussed.

\section{Tropical Gibbs' distribution and free energy}

So we will consider macroscopic systems in equilibrium and will study
their limiting behavior as (formally) $k\rightarrow0$. Introducting
the energy level ``entropy'' $S_{n}=k\ln g_{n}$ and assuming that $S_{n}$ are finite, one has the following form of partition function 

\begin{equation}
{\displaystyle Z=\sum_{n\geq1}\exp\left[\frac{1}{k}\left(S_{n}-\frac{E_{n}}{T}\right)\right]=\sum_{n\geq1}\exp\left(-\frac{F_{n}}{kT}\right)}\label{eq: partition function}
\end{equation}
where $F_{n}\equiv E_{n}-TS_{n}$ is the ``energy-level'' free energy
and energies $E_{n}$ are ordered as $0<E_{1}<E_{2}<...$ . One observes
that the degeneracies ${\displaystyle g_{n}=\exp\frac{S_{n}}{k}}$ 
with finite $S_{n}>0$ and Boltzmann weights ${\displaystyle \exp\left(-\frac{E_{n}}{kT}\right)}$
behave quite differently as $k\rightarrow0$. So in the tropical limit
we will have sort of Bergmann's logarithmic limit set \cite{35}.  

Tropical limit of probability $w_{n}$, in general, is naturally associated
with its singular behavior of the form $w_{n}={\displaystyle \tilde{w}_{n}\cdot\exp\frac{w_{n,tr}}{\varepsilon}}$
with small positive parameter $\varepsilon$, $0<\tilde{w}_{n}\leq1$
and $w_{n,tr}={\displaystyle \lim_{\varepsilon\rightarrow0} \left(\varepsilon \ln w_{n}\right)}$. 
Tropical probability $w_{n,tr}$ varies in the interval $(-\infty,0]$. 
The interval $0<w_{n}\leq1$ collapses into $\{0\}$ while exponentially
small usual probabilities $w_{n}$ are represented by the whole
semi-line $(-\infty,0)$ for $w_{n,tr}$ and numbers $\tilde{w}_{n}$. The meaning of the quantities $w_{n,tr}$ and $\hat{w}_{n}$ is clarified
by the formula ${\displaystyle \ln w_{n}=\frac{w_{n,tr}}{\varepsilon}+\ln\tilde{w}_{n}+...}$ . 
So singular behavior under consideration is characterized
by a simple pole behavior of $\ln w_{n}$ as a function of the small
parameter $\varepsilon$: $w_{n,tr}$ is the residue at this pole
while $\ln\tilde{w}_{n,tr}$ is the first regular nondominant term. 
In generic regular case it is sufficient to consider the dominant
pole term and, hence, the tropical probability $w_{n,tr}$. Contribution
of nondominant term $\ln\tilde{w}_{n}$ becomes crucial, as we shall see, in the singular situations when limit $\varepsilon\rightarrow0$ ceases to be uniquely
defined.

Under the assumption that all $F_{n}$ are distinct the tropical limit of Gibbs'
probabilities (\ref{eq: Gibbs' distribution}) is given
by ($\varepsilon= k$) 

\begin{equation}
{\displaystyle w_{n,tr}=-\frac{E_{n}}{T}-\max\left\{ -\frac{F_{1}}{T},-\frac{F_{2}}{T},...\right\} =-S_{n}-\frac{F_{n}}{T}+\min\left\{ \frac{F_{1}}{T},\frac{F_{2}}{T},...\right\} }.\label{eq: tropical Gibbs probabilities, k}
\end{equation}
Denoting ${\displaystyle \left(\frac{F}{T}\right)_{min}:=\min\left\{ \frac{F_{1}}{T},\frac{F_{2}}{T},...\right\} }$,
one gets 

\begin{equation}
{\displaystyle w_{n,tr}=-S_{n}-\frac{F_{n}}{T}+\left(\frac{F}{T}\right)_{min}}.\label{eq: tropical Gibbs probabilities, k bis}
\end{equation}
Normalization condition for these tropical probabilities is the limit $k \rightarrow 0$ of the condition $\displaystyle{\sum_{n} g_{n}\cdot w_{n}=1}$ and it is given by 
\begin{equation}
{\displaystyle \sum_{n} \oplus \left(S_{n} + w_{n,tr} \right)=0}.\label{eq: normalization condition for w_{n}}
\end{equation}
In particular, for $n=n_{0}$ such that ${\displaystyle \frac{F_{n_{0}}}{T}=\left(\frac{F}{T}\right)_{min}}$,
one has 

\begin{equation}
{\displaystyle {\displaystyle w_{n_{0},tr}=-S_{n_{0}}}.}\label{eq: tropical Gibbs probability at n_0}
\end{equation}

So, the entropies $S_{n}$ are, in fact, the tropical Gibbs' probabilities
to find the system in certain state with energy $E_{n}$. Probability
$W_{n}$ for the system to have energy $E_{n}$ at small $k$ and
$T>0$ is equal to ${\displaystyle W_{n}=g_{n}\exp \frac{w_{n,tr}}{k}=\exp \frac{W_{n,tr}}{k}}$ and, hence, tropical probability $W_{n,tr}$ for the system to have energy $E_{n}$ is equal to 
\begin{equation}
{\displaystyle W_{n,tr}=\frac{F_{tr}-F_{n}}{T}}.\label{eq: Tropical probability for energy levels}
\end{equation}
These tropical probabilities obey the normalization condition  $\displaystyle \sum_{n} \oplus W_{n,tr}=\max \{W_{n,tr}\}=0$. Also in the limit $k\rightarrow 0$ for usual probabilities one gets 
$W_{n_{0}}=1$ and $W_{n\neq n_{0}}=0$ and the tropical energy
$E_{tr}$ of the system is 
\begin{equation}
{E_{tr}={\displaystyle \lim_{k \rightarrow 0} \left( \sum_{n\geq1} W_{n}	E_{n} \right)=E_{n_{0}}}} \label{eq: tropical energy}.
\end{equation}
The tropical Gibbs' distribution provides us with the fine description
of the energy levels. 

Tropical limit of the free energy (\ref{eq: free energy}) is given
by 

\begin{equation}
{\displaystyle F_{tr}(T)=-T\max\left\{-\frac{F_{1}}{T},-\frac{F_{2}}{T},...\right\}=T\min\left\{\frac{F_{1}}{T},\frac{F_{2}}{T},...\right\}}\label{eq: tropical free energy 1}
\end{equation}
or 

\begin{equation}
{\displaystyle -\frac{F_{tr}(T)}{T}=-\sum_{n\geq1}\oplus\left\{-\frac{F_{n}}{T}\right\}},\label{eq: tropical free energy 2}
\end{equation}
assuming that ${\displaystyle \max\{-\frac{F_{n}}{T},n\geq1\}}$ exists. Thus, the tropical free energy is the additive tropical sum of the
free energies $F_{n}$ of energy levels. 

Tropical free energy $F_{tr}$ is, in general, a piecewise linear
function of temperature $T$. For instance, for positive $T$ it is 

\begin{equation}
{\displaystyle F_{tr}=\min\{E_{1}-TS_{1},E_{2}-TS_{2},...\}}.\label{eq: tropical free energy, T>0}
\end{equation}
In this case the tropical limit $S_{tr}$ of the entropy defined
by the standard formula $S_{tr}={\displaystyle -\frac{\partial F_{tr}}{\partial T}}$
is equal to $S_{tr}=S_{n_{0}}$ where $F_{n_{0}}=F_{min}$. With
such a definition of $F_{tr}$ and $S_{tr}$ one has 

\begin{equation}
{\displaystyle dF_{tr}=-S_{tr}dT-p_{tr}dV}\label{eq: tropical thermodynamic law}
\end{equation}
where $p_{tr}={\displaystyle -\left(\frac{\partial F_{tr}}{\partial V}\right)_{T}}$
and $F_{tr}=E_{tr}-TS_{tr}$. 

The same tropical entropy is obtained as the limit $k\rightarrow0$
of the standard formula $S=-k{\displaystyle \overline{\ln w_{n}}}$.
Indeed, at $T>0$ the Gibbs' probabilities are 

\begin{equation}
{\displaystyle w_{n,tr}=-S_{n}+\frac{F_{tr}-F_{n}}{T}}\label{eq: tropical Gibbs probabilities}
\end{equation}
and 

\begin{equation}
{\displaystyle w_{n_{0},tr}=-S_{tr}}.\label{eq: tropical Gibbs probability and tropical entropy}
\end{equation}
Using these formulas, one gets 

${\displaystyle -\lim_{k\rightarrow0}(k\overline{\ln w_{n}})=-\lim_{k\rightarrow0} \left(k\sum_{n\geq1}g_{n}w_{n}\ln w_{n}\right)}=$

${\displaystyle =-\lim_{k\rightarrow0}\left(k\sum_{n\geq1}\exp\left(\frac{S_{n}}{k}\right)\cdot\left[\exp\left(\frac{w_{n,tr}}{k}+\mathcal{O}(k)\right)\right]\cdot\left[-\frac{S_{n}}{k}-\frac{F_{n}-F_{tr}}{kT}+\mathcal{O}(k)\right]\right)=}$ 

\begin{equation}
{\displaystyle =S_{n_{0}}=S_{tr}.}\label{eq: tropical entropy regular case}
\end{equation}

Tropical entropy is a piecewise constant function of temperature $T$.
At the limit $T\rightarrow0$ one has $S_{tr}=S_{n_{0}}(T\rightarrow +0)=S_{1}$.
So $S_{1}$ is the residual entropy of the system at absolute zero.
Tropical entropy $S_{tr}$ is a constant and hence the specific heat
$C_{V,tr}=T\left({\displaystyle \frac{\partial S_{tr}}{\partial T}}\right)_{V}=0$.
At the limit $T\rightarrow -0$, $S_{tr}=S_{n^{*}}$ , where $S_{n^{*}}$
is the entropy of the level with largest energy. 

Formulae (\ref{eq: normalization condition for w_{n}}, \ref{eq: tropical Gibbs probability at n_0}, \ref{eq: Tropical probability for energy levels}) have a simple probabilistic interpretation. Indeed, the tropical probability $W_{n,tr}$ for the system to have energy $E_{n}$ is given by (\ref{eq: Tropical probability for energy levels}) where $F_{tr}=F_{min}=F_{n_{0}}$. Thus, the system at temperature $T(>0)$ has such energy $E_{n}$ for which the tropical probability $W_{n,tr}$ is maximal, i.e. zero. This law of tropical probability maximum is a clear manifestation of the relevance of the limit $k \rightarrow 0$. 

Finally, we note that in the case of all $S_{n}=0$ one has the tropical limit for the system with finite degeneracies for which $F_{tr}=E_{1}$.  

\section{Singularity in the tropical limit}

The formulae and results presented in the previous section are valid
in generic situation when all $F_{n}$ are distinct. 

Singularity (nonuniqueness) of the tropical limit arises in the case
when $F_{min}$ is attained on two or more $F_{n}$ (see e.g. \cite{9,10,12,13}).
In such a situation the tropical free energy $F_{tr}$ is nondifferentiable
as a function of temperature $T$ at certain $T=T^{*}$. 
\\
How other tropical quantities behave at these temperatures? 
Let $F_{n_{0}}$ and $F_{n_{0}+1}$ be two successive minima of
$\{F_{n}\}$ with $S_{n_{0}}\neq S_{n_{0}+1}$. At $F_{n_{0}}=F_{min}$
tropical entropy is $S_{n_{0}}$. For $F_{min}=F_{n_{0}+1}$ one has
$S_{tr}=S_{n_{0}+1}$. \\
Values of $F_{n_{0}}$ and $F_{n_{0}+1}$ coincide at the temperature
($F_{min}(T^{*})=F_{n_{0}}(T^{*})=F_{n_{0}+1}(T^{*})$) 

\begin{equation}
{\displaystyle T^{*}=\frac{E_{n_{0}}-E_{n_{0}+1}}{S_{n_{0}}-S_{n_{0}+1}}}.\label{eq: critical temperature}
\end{equation}

Sign of $T^{*}$ coincides with $\mbox{sign}(S_{n_{0}+1}-S_{n_{0}})$.
If $S_{n_{0}+1}>S_{n_{0}}$ then $F_{n_{0}}<F_{n_{0}+1}$ at $T<T^{*}$ and
$F_{n_{0}+1}<F_{n_{0}}$ at $T>T^{*}$. So at $T<T^{*}$ the tropical
entropy is equal to $S_{n_{0}}$ while at $T>T^{*}$ one has $S_{tr}=S_{n_{0}+1}$.
Thus, the tropical entropy jumps when the temperature $T$ passes
the point $T^{*}$. 
This jump of entropy $S_{tr}$ at $T=T^{*}$ is the tropical trace
of the entropy drop phenomenon discussed for spin glasses in \cite{28,30,33}. 
Using the formula (\ref{eq: Tropical probability for energy levels}) for the difference of tropical probability $W_{n_{0},tr}(T)$ and $W_{n_{0}+1,tr}(T)$, one gets
\begin{equation}
{\displaystyle \Delta W_{tr}=W_{n_{0},tr}(T) - W_{n_{0}+1,tr}(T)=\frac{F_{n_{0}+1}-F_{n_{0}}}{T}=\Delta S_{tr} \cdot \left( \frac{T^{*}}{T}-1 \right)}.\label{eq: Delta tropical probabilities}
\end{equation}
So $\Delta W_{tr}$ behaves as $\sim \left( T^{*}-T \right)$ near the singular temperature $T^{*}$. 

In order to calculate probabilities and entropy at singular point
$T=T^{*}$ one should take into account not only dominant terms but
also first regular nondominant terms. 
Indeed in the case, where $F_{min}$ is attained precisely on $F_{n_{0}}$
and $F_{n_{0}+1}$, from the Gibbs' distribution (\ref{eq: Gibbs' distribution})
at small $k$ one gets 

\begin{equation}
{\displaystyle \ln w_{n}=-\frac{S_{n}}{k}-\frac{F_{n}(T^{*})-F_{n_{0}}(T^{*})}{kT^{*}}-\ln2+\mathcal{O}(k)}.\label{eq: tropical Gibbs distribution + first order}
\end{equation}
In particular 

\begin{equation}
{\displaystyle \begin{array}{c}
{\displaystyle \ln w_{n_{0}}=-\frac{S_{n_{0}}}{k}-\ln2+\mathcal{O}(k)},\\
{\displaystyle \ln w_{n_{0}+1}=-\frac{S_{n_{0}+1}}{k}-\ln2+\mathcal{O}(k)},\\
{\displaystyle W_{n_{0},tr}=W_{n_{0}+1,tr}=-k \cdot \ln 2}. 
\end{array}}\label{eq: tropical Gibbs distribution + first order, special cases}
\end{equation}
Hence, at $T=T^{*}$ and small $k$ 

\begin{equation}
{\displaystyle \begin{array}{c}
{\displaystyle w_{n_{0}}=\frac{1}{2}\exp\left(-\frac{S_{n_{0}}}{k}\right)},\\
{\displaystyle w_{n_{0}+1}=\frac{1}{2}\exp\left(-\frac{S_{n_{0}+1}}{k}\right)}.
\end{array}}\label{eq: tropical Gibbs distribution + first order, special cases-1}
\end{equation}
Using definition of tropical entropy, one readily obtains 

\begin{equation}
{\displaystyle S_{tr}(T^{*})=-\lim_{k\rightarrow0}(k\overline{\ln w(T^{*})})=\frac{1}{2}(S_{n_{0}}+S_{n_{0}+1})}.\label{eq: tropical Gibbs distribution, entropy}
\end{equation}
One gets the same result by direct calculation, namely, \\
${\displaystyle S_{tr}(T^{*})=-\lim_{k\rightarrow0}(k\overline{\ln w(T^{*})})=}$ \\
${\displaystyle =-\lim_{k\rightarrow0}\left[k\frac{{\displaystyle \exp\left(-\frac{F_{n_{0}}}{kT}\right)}}{{\displaystyle 2\exp\left(-\frac{F_{n_{0}}}{kT}\right)+\sum_{n\neq n_{0},n_{0}+1}\exp\left(-\frac{F_{n}}{kT}\right)}}\left(-\frac{S_{n_{0}}}{k}-\frac{S_{n_{0}+1}}{k}+\mathcal{O}(k^{0})\right)\right]-}$ 
\\
${\displaystyle {\displaystyle -\lim_{k \rightarrow0}\left[k\sum_{m\neq n_{0},n_{0}+1}\frac{{\displaystyle \exp\frac{F_{n_{0}}-F_{m}}{kT}}}{{\displaystyle \left(2+\sum_{n\neq n_{0},n_{0}+1}\exp\frac{F_{n_{0}}-F_{n}}{kT}\right)}} \left(-\frac{S_{m}}{k}-\frac{F_{m}-F_{n_{0}}}{kT}+\mathcal{O}(k^{0})\right)\right]=}}$

\begin{equation}
{\displaystyle =\frac{1}{2}(S_{n_{0}}+S_{n_{0}+1}).}\label{eq: tropical Gibbs distribution, entropy bis}
\end{equation}

Note that at $T=T^{*}$ the usual probabilities $W_{n_{0}}=g_{n_{0}}\cdot w_{n_{0}}$
and $W_{n_{0}+1}=g_{n_{0}+1}\cdot w_{n_{0}+1}$ for the system to
have energies $E_{n_{0}}$ and $E_{n_{0}+1}$, respectively, are equal
${\displaystyle W_{n_{0}}=W_{n_{0}+1}}$\\${\displaystyle = \frac{1}{2}}$. One also has $E_{tr}(T^{*})={\displaystyle \frac{1}{2}(E_{n_{0}}+E_{n_{0}+1})}$. 
Note that in the generic case one has $\ln1$ instead of $\ln2$ in
(\ref{eq: tropical Gibbs distribution + first order}) and, hence,
the formula (\ref{eq: tropical Gibbs probability and tropical entropy}). 
If $F_{min}$ would be attained on $m$ $F_{n}$'s, one would have
${\displaystyle \ln w_{n_{0}}=-\frac{S_{n_{0}}}{k}-\ln m+\mathcal{O}(k)}$. 

In more details the importance of nondominant terms in singular points
and their relevance to the deformation of idempotent semiring will
be discussed elsewhere.

\section{Systems with finite number of energy levels}

We begin with the simplest non-trivial case of two level systems and
$S_{2}>S_{1}$. 
For the systems with finite number of energy levels there is no constraint
on the sign of temperature (see \cite{22}). Thus the tropical free energy at $T>0$ is 

\begin{equation}
{\displaystyle F_{tr}(T)=\min\{E_{1}-TS_{1},E_{2}-TS_{2}\}}\label{eq: 2 levels, tropical free energy at T>0}
\end{equation}
and 

\begin{equation}
{\displaystyle F_{tr}=\max\{E_{1}-S_{1}T,E_{2}-S_{2}T\}}\label{eq: 2 levels, tropical free energy at T<0}
\end{equation}
for $T<0$. 

Transition temperature $T^{*}={\displaystyle \frac{E_{2}-E_{1}}{S_{2}-S_{1}}}>0$. 
At $0<T<T^{*}$ one has 

\begin{equation}
{\displaystyle F_{tr}(T)=E_{1}-S_{1}T}\label{eq: 2 levels, Ftr, 0<T<T*}
\end{equation}
and tropical energy and entropy are $E_{tr}=E_{1}$ and $S_{tr}=S_{1}$. 

At $T>T^{*}$ and $T<0$ 

\begin{equation}
{\displaystyle F_{tr}(T)=E_{2}-S_{2}T}\label{eq: 2 levels, Ftr, T>T* or T<0}
\end{equation}
and $E_{tr}=E_{2}$, $S_{tr}=S_{2}$. 

\begin{figure}[tph]
\centering
\protect\includegraphics[scale=0.6]{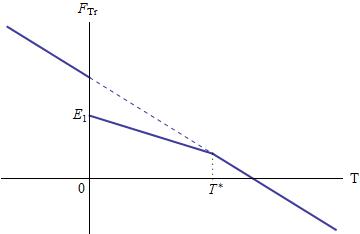}
 \caption{$F_{tr}$ in two level case, $S_{2}>S_{1}$.} 
\label{fig: Figure 1}
\end{figure}

The graph of $F_{tr}(T)$ (at ${\displaystyle \frac{E_{2}}{E_{1}}<\frac{S_{2}}{S_{1}}}$) is presented in figure (\ref{fig: Figure 1}). 

The residual entropy is equal to $S_{tr}(T\rightarrow +0)=S_{1}$,
while $S_{tr}(T\rightarrow -0)=S_{2}$. The jump of $S_{tr}$ at
$T=T^{*}$ is $S_{2}-S_{1}$ and $S_{tr}(T^{*})={\displaystyle \frac{1}{2}(S_{1}+S_{2})}$,
$E_{tr}(T^{*})={\displaystyle \frac{1}{2}(E_{1}+E_{2})}$. \\
At $T<T^{*}$ the tropical probability $W_{1,tr}$ to have energy $E_{1}$ is equal to $W_{1,tr}=0$ while the probability $W_{2,tr}$ to have energy $E_{2}$ is $\displaystyle W_{2,tr}=(S_{2}-S_{1})\cdot \left( 1- \frac{T^{*}}{T}\right)<0$. At $T>T^{*}$ the situation is opposite, namely, $W_{2,tr}=0$, $\displaystyle W_{1,tr}=(S_{1}-S_{2})\cdot \left( 1- \frac{T^{*}}{T}\right) <0$. 

\begin{figure}[tph]
\centering
\protect\includegraphics[scale=0.5]{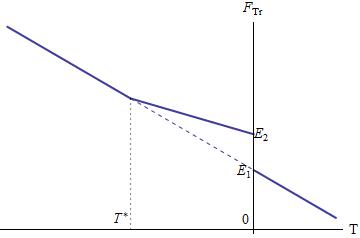}
\caption{$F_{tr}$ in two level case, $S_{1}>S_{2}$.} 
\label{fig: Figure 2}
\end{figure}

In the case $S_{1}>S_{2}$ the transition temperature $T^{*}$ is
negative and the graph of $F_{tr}(T)$ is given on the figure (\ref{fig: Figure 2}). 

So $E_{tr}(T\rightarrow +0)=E_{1}$ and $S_{tr}(T\rightarrow +0)=S_{1}$
again, $S_{tr}(T\rightarrow -0)=S_{2}$ and $S_{tr}(T^{*})={\displaystyle \frac{1}{2}(S_{1}+S_{2})}$,
$E_{tr}(T^{*})={\displaystyle \frac{1}{2}(E_{1}+E_{2})}$. 
Since negative temperatures are higher than positive one, in this
case transition takes place in ``higher'' temperature. 

\begin{figure}[tph]
\centering
\protect\includegraphics[scale=0.5]{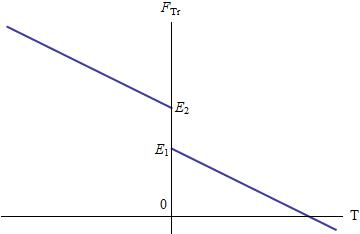}
\caption{$F_{tr}$ in two level case, $S_{1}=S_{2}$.} 
\label{fig: Figure 3}
\end{figure}

Finally, if $S_{1}=S_{2}$, one has (see figure (\ref{fig: Figure 3})) 
$T^{*}=+\infty$, $E_{tr}(T>0)=E_{1}$, $S_{tr}(T>0)=S_{1}$,
$E_{tr}(T<0)=E_{2}$, $S_{tr}(T<0)=S_{2}$. 

For the system with three energy levels $E_{1}<E_{2}<E_{3}$ we will
consider here only three characteristic cases. 
\begin{enumerate}
\item $S_{1}<S_{2}<S_{3}$. At $T>0$, ${\displaystyle F_{tr}(T)=\min\{E_{1}-S_{1}T,E_{2}-S_{2}T,E_{3}-S_{3}T\}}.$
\\
Singular (transition) values of $T$ are \\
\begin{equation}
{\displaystyle T_{ik}^{*}=\frac{E_{i}-E_{k}}{S_{i}-S_{k}}>0},\,\,\, i,k=1,2,3,\, i\neq k.\label{eq: transition temperatures}
\end{equation}
\\
They are not independent since \\
\begin{equation}
{\displaystyle (S_{1}-S_{2})T_{12}^{*}+(S_{2}-S_{3})T_{23}^{*}+(S_{3}-S_{1})T_{31}^{*}=0}.\label{eq: dependence relation transition temperatures}
\end{equation}
\\
As the consequence, $T_{12}^{*}-T_{23}^{*}$ and $T_{13}^{*}-T_{23}^{*}$
have the same sign. \\
In the case $T_{23}^{*}>T_{12}^{*}$ and $T_{13}^{*}>T_{12}^{*}$
the graph of $F_{tr}$ is given on the figure (\ref{fig: Figure 4}) \\
\begin{figure}[tph]
\centering

\protect\includegraphics[scale=0.5]{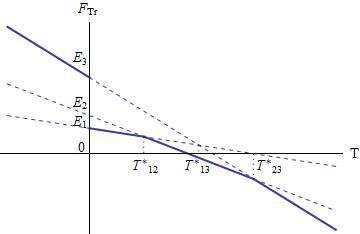}
\caption{$F_{tr}$ in three level case, $S_{1}<S_{2}<S_{3}$, $T_{23}^{*}>T_{13}^{*}>T_{12}^{*}$.} 
\label{fig: Figure 4}
\end{figure}
 \\
So at $0<T<T_{12}^{*}$, $F_{tr}=E_{1}-S_{1}T$, $E_{tr}=E_{1}$,
$S_{tr}=S_{1}$. For $T_{12}^{*}<T<T_{23}^{*}$ one has $F_{tr}=E_{2}-S_{2}T$,
$E_{tr}=E_{2}$, $S_{tr}=S_{2}$; at $T>T_{23}^{*}$ and $T<0$ it's
$F_{tr}=E_{3}-S_{3}T$, $E_{tr}=E_{3}$, $S_{tr}=S_{3}$. 
\item If, instead, $T_{23}^{*}<T_{13}^{*}<T_{12}^{*}$ one has quite different
behavior of the system (figure \ref{fig: Figure 5}).  
\begin{figure}[tph]
\centering
\protect\includegraphics[scale=0.5]{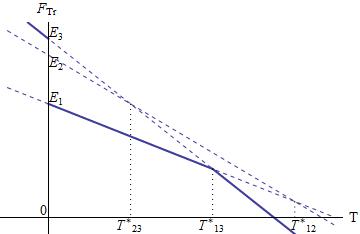}
\caption{$F_{tr}$ in three level case, $S_{1}<S_{2}<S_{3}$, $T_{23}^{*}<T_{13}^{*}<T_{12}^{*}$.} 
\label{fig: Figure 5}
\end{figure}
So at $0<T<T_{13}^{*}$, $F_{tr}=E_{1}-S_{1}T$, $E_{tr}=E_{1}$,
$S_{tr}=S_{1}$, while for $T>T_{13}^{*}$, $F_{tr}=E_{3}-S_{3}T$,
$E_{tr}=E_{3}$, $S_{tr}=S_{3}$. \\
Thus depending on parameter $E_{i}$, $S_{i}$ one may have two singular
values of temperature or only one. In the first case, with increase
of temperature from $T=0$, the system in the tropical limit first
jumps from the macroscopic state of the system with energy $E_{1}$ and entropy $S_{1}$ to
the state with energy $E_{2}$ and entropy $S_{2}$ at temperature
$T_{12}^{*}$, and then at temperature $T_{23}^{*}$ it jumps from
$E_{2}$ and $S_{2}$ to $E_{3}$ and $S_{3}$ respectively. \\
In the second case, the system jumps only once at temperature $T_{13}^{*}$
from the state with $E_{1}$, $S_{1}$ to the state with $E_{3}$,
$S_{3}$. \\
At $S_{1}>S_{2}>S_{3}$ one has similar behavior at negative $T$. 
\item Our third example corresponds to $S_{3}<S_{1}<S_{2}$. In this case
$T_{23}^{*}<0$ and $F_{tr}$ is given in figure (\ref{fig: Figure 6}).
\\
\begin{figure}[tph]
\centering
\protect\includegraphics[scale=0.5]{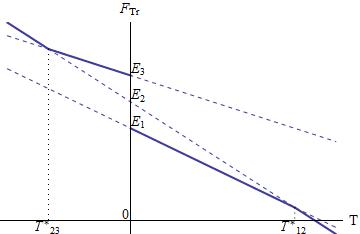}
\caption{$F_{tr}$ in three level case, $S_{3}<S_{1}<S_{2}$.} 
\label{fig: Figure 6}
\end{figure}
\\
So at $0<T<T_{12}^{*}$, $E_{tr}=E_{1}$, $S_{tr}=S_{1}$. At $T_{12}^{*}<T<+\infty$
one has $E_{tr}=E_{2}$, $S_{tr}=S_{2}$. At $-\infty<T<T_{23}^{*}$,
$F_{tr}=E_{2}-S_{2}T$, $E_{tr}=E_{2}$, $S_{tr}=S_{2}$, while at
$0>T>T_{23}^{*}$, $E_{tr}=E_{3}$, $S_{tr}=S_{3}$. 
\end{enumerate}

In all above three cases the system has at most two singular (transition)
temperatures. One can show that the same is true in all other cases. 

For the system with $n$ energy levels there is number of different
cases. In principle there are ${\displaystyle \frac{n\cdot(n-1)}{2}}$
singular values of temperature $T_{ij}^{*}={\displaystyle \frac{E_{i}-E_{j}}{S_{i}-S_{j}}}=T_{ji}^{*}$,
for $i,\, j=1,...,n$, $i\neq j$. Due to the identities 
\begin{equation}
{\displaystyle {\displaystyle (S_{i}-S_{k})T_{ik}^{*}+(S_{k}-S_{l})T_{kl}^{*}+(S_{l}-S_{i})T_{li}^{*}=0},\, i\neq k\neq l\neq i,\, i,\, k,\, l=1,...,n},\label{eq: dependence relations transition temperatures, n levels}
\end{equation}
there is a number of constraints on different $T_{ik}^{*}$. It is
not difficult to show that the tropical free energy $F_{tr}(T)$ may
have at most $n$ linear pieces and hence $n-1$ transition temperatures
$T^{*}$. In all situations, $E_{tr}(T\rightarrow +0)=E_{1}$, $S_{tr}(T\rightarrow +0)=S_{1}$
and $E_{tr}(T\rightarrow -0)=E_{n}$, $S_{tr}(T\rightarrow -0)=S_{n}$. 

At $T>0$ the thermodynamical energy $E_{tr}$ of
the system with $n$ energy levels may assume one, two, up to $n$
different values depending on relations between $S_{1}$, $S_{2}$,
..., $S_{n}$. 

We would like to note that systems with finite number of energy levels
appear also in the study of frustrated systems with different ``micro-basins''
of macroscopic systems characterized by energies $E_{n}$. For such
type of systems the transition temperatures $T^{*}$ considered above
represent the tropical limit of the points of first order phase transitions
with latent heats $q_{ik}=T_{ik}^{*}(S_{i}-S_{k})=E_{i}-E_{k}$.

\section{Systems with infinitely many energy levels and limiting temperatures}

Tropical limit for systems with countable number of energy levels
is defined in the same manner as in the finite case. The
tropical free energy $F_{tr}$ is an infinite tropical sum 

\begin{equation}
F_{tr}(T)={\displaystyle -T \sum_{n}\oplus \left(-\frac{F_{n}}{T} \right)}\label{eq: Tropical free energy}
\end{equation}
viewed as the formal limit of the standard tropical sum when the number
of terms goes to infinity. 

For the systems with the energy spectrum bounded from below, the sum
in (\ref{eq: Tropical free energy}) is performed over all $n\geq1$,
with $0<E_{1}<E_{2}<...<E_{n}<...$ . If the energy spectrum of the system
is unbounded from below, then the sum is over all integers $-\infty<n<\infty$
and energies ordered as $E_{n}<E_{n+1}$. 

Existence or nonexistence of $F_{tr}(T)$ for certain temperatures,
i.e. convergence or divergence of the infinite sum in (\ref{eq: Tropical free energy}),
is the key novel feature of infinite case in comparison with the finite
one. This is the tropical $k\rightarrow0$ trace of the existence
of limiting temperatures for usual macroscopic systems, i.e. existence
of intervals of temperature for which system cannot stay in equilibrium
(see \cite{22,36,37}). 

Let us begin with a system with the energy spectrum bounded from
below and unbounded from above. 
At $T>0$ one has ${\displaystyle F_{tr}(T)=-T\max\left\{-\frac{F_{1}}{T},-\frac{F_{2}}{T},...,\frac{F_{n}}{T},...\right\}}$$=\min\{F_{1},F_{2},F_{3},...\}$. 
In the case when the minimum of $\{F_{1},F_{2},...\}$ exists, i.e.
it is finite, one has the behavior which is just the limit of that
considered in the previous section. 

Singularity of $F_{tr}(T)$ may occur for such a temperature $T_{S}$
when, beginning from certain $F_{n_{0}}(T)$ all $F_{n_{0}+k}(T)>F_{n_{0}+k+1}(T)$
(at $T>T_{S}$ and $k=0,1,2,...$) and, hence, $\min\{F_{1},F_{2},...\}$
does not exist. 
It is achieved, for instance, in the case when $S_{n_{0}+k+1}>S_{n_{0}+k}>0$, ${\displaystyle \frac{E_{n_{0}+k+1}}{S_{n_{0}+k+1}}\leq\frac{E_{n_{0}+k}}{S_{n_{0}+k}}},$
$k=0,1,2,...$ and set of differencies $S_{n_{0}+k+1}-S_{n_{0}+k}$ has positive lower bound. Indeed, it is easy to see, that at $T>T_{S+}={\displaystyle \frac{E_{n_{0}}}{S_{n_{0}}}}$
one has 

${\displaystyle F_{n_{0}}(T)=S_{n_{0}}(T_{S+}-T)>F_{n_{0}+1}(T)=S_{n_{0}+1}\left(\frac{E_{n_{0}+1}}{S_{n_{0}+1}}-T\right)>...}$

$\displaystyle{...F_{n_{0}+k}=S_{n_{0}+k}\left(\frac{E_{n_{0}+k}}{S_{n_{0}+k}}-T\right)>...}$,
$k=2,3,...$ . 

Thus, for $T>T_{S+}$, this sequence has no lower limit, so $\min\{F_{n}\}$ is
not bounded from below, infinite tropical sum diverges and $F_{tr}(T)$
does not exist. So our system can be in equilibrium only for temperatures
$T$ in the interval $0\leq T<T_{S+}$. 
This phenomenon has a simple probabilistic interpretation similar to that of full nontropical case (see e.g. \cite{36,37}). Indeed, for "forbidden" values of temperature, $F_{tr}=F_{min}\rightarrow -\infty$. Even if one works in $\mathbb{R}_{\max}=(\mathbb{R}\cup\{-\infty\},\max,+)$, and hence $-\infty \in \mathbb{R}_{\max}$, the formula (\ref{eq: Tropical probability for energy levels}) for $W_{n,tr}$ does not define any distribution of probabilities obeying the tropical normalization condition and the system cannot stay in equilibrium.   

In the particular case $n_{0}=1$ and ${\displaystyle \frac{E_{n}}{S_{n}}=a}$,
i.e. $S_{n}={\displaystyle \frac{E_{n}}{a}}$, $n=1,2...$, $F_{n}=E_{n}{\displaystyle \left(1-\frac{T}{a}\right)}$
the limiting temperature is $T_{S+}=a$. Such limiting temperature
is (not surprisingly) the same as in the corresponding full nontropical
case (see \cite{37}). In the
tropical limit the system with temperature $T<T_{S+}={\displaystyle \frac{E_{n_{0}}}{S_{k_{0}}}}$
has the same properties as the system with $n_{0}$ energy levels,
i.e. it may have at most $n_{0}$ values of energy and entropy. In
particular, in the case $n_{0}=1$, e.g. ${\displaystyle S_{n}={\displaystyle \frac{E_{n}}{a}}}$,
the system has energy $E_{1}$ and entropy $S_{1}$ for all $T<T_{S+}$. 

Different limits for temperature arise in the case when all $S_{n_{0}+k}<0$,
$k=0,1,2,...$ for some $n_{0}$. Indeed, if $S_{n_{0}+k+1}<S_{n_{0}+k}<0$,
and ${\displaystyle \frac{S_{n_{0}+k+1}}{E_{n_{0}+k+1}}\geq\frac{S_{n_{0}+k}}{E_{n_{0}+k}}}$,
$k=0,1,2,...$, then at ${\displaystyle \frac{E_{n_{0}}}{S_{n_{0}}}<T<0}$ 
one has  
${\displaystyle -\frac{F_{n_{0}}}{T}=E_{n_{0}}\left(\frac{S_{n_{0}}}{E_{n_{0}}}-\frac{1}{T}\right)<-\frac{F_{n_{0}+1}}{T}=E_{n_{0}+1}\left(\frac{S_{n_{0}+1}}{E_{n_{0}+1}}-\frac{1}{T}\right)<...}$\\
$\displaystyle{...<-\frac{F_{n_{0}+k}}{T}<...}$ . So, $\max{\displaystyle \left\{-\frac{F_{n}}{T}\right\}}$ does not exist and
$F_{tr}(T)$ is not defined. Thus in this case one has the interval
${\displaystyle -\frac{E_{n_{0}}}{|S_{n_{0}}|}<T<0}$ of ``forbidden''
temperatures. 
In the particular case $S_{n}={\displaystyle -\frac{E_{n}}{a}}$,
$n=1,2,...$ ($a>0$), this interval is ${\displaystyle -a<T<0}$
and it represents the tropical limit of the situation discussed in
\cite{37}. 

Systems with unbounded from below energy spectrum may have lower limit
for temperature (see e.g. \cite{38}).
In the tropical limit one has ($T>0$) 

\begin{equation}
F_{tr}=-T{\displaystyle \sum_{-\infty<n<\infty}\oplus\left(-\frac{F_{n}}{T}\right)=\min\{...,F_{-2},F_{-1},F_{0},F_{1},F_{2},...\}}\label{eq: tropical free energy unbounded from below spectrum}
\end{equation}
where $E_{n}<0$ at $n<0$. The tropical sum in (\ref{eq: tropical free energy unbounded from below spectrum})
might diverge due to the negative terms $F_{-n}=E_{-n}-S_{-n}T$ ($n=1,2,...$).
Indeed, let there exist certain $m_{0}>0$ such that all $S_{-m_{0}-k}$
($k=0,1,2,...$) are negative, $|S_{-m_{0}-k-1}|>|S_{-m_{0}-k}|$ ($k=0,1,2,...$)
and ${\displaystyle \frac{E_{-m_{0}-k}}{S_{-m_{0}-k}}\leq\frac{E_{-m_{0}-k-1}}{S_{-m_{0}-k-1}}}$,
$k=0,1,...$ . Then at $T<T_{S-}:={\displaystyle \frac{E_{-m_{0}}}{S_{-m_{0}}}}$
one has 
\begin{align}
F_{-m_{0}}=|S_{-m_{0}}|\cdot(T-T_{S-})>|S_{-m_{0}-1}|\cdot \left(T-\frac{E_{-m_{0}-1}}{S_{-m_{0}-1}}\right)>... \nonumber \\ 
...>|S_{-m_{0}-k}|\cdot \left(T-\frac{E_{-m_{0}-k}}{S_{-m_{0}-k}}\right)>...\,. \label{eq: chain for unbounded from below spectrum}
\end{align}

So the sequence $F_{-m_{0}}$, $F_{-m_{0}-1}$, $F_{-m_{0}-2}$,...
is strictly decreasing and it can be unbounded from below at $T<T_{S-}$.
In such a case $F_{tr}(T)$ does not exist at $T<T_{S-}$ and $T_{S-}$ is
the lower limiting temperature. 

Combining the above case and that considered the first in this section,
one gets the system with unbounded both from below and above energy
spectrum which can stay in equilibrium at the temperature belonging
to the finite interval $T_{S-}<T<T_{S+}$ with 
\medskip
$T_{S+}={\displaystyle \frac{E_{n_{0}}}{S_{n_{0}}}}$ and $T_{S-}={\displaystyle \frac{E_{-m_{0}}}{S_{-m_{0}}}}$. 

More complicated cases with intervals of allowed (or forbidden) temperatures
as well as the behavior of such systems at $T=T_{S}$ will be discussed
elsewhere. 

\bigskip

{\bf Acknowledgments.} The second author (B.K.) was partially supported
by the PRIN 2010/2011 grant 2010JJ4KBA\_003. 


\cleardoublepage
\addcontentsline{toc}{section}{References}


\end{document}